\documentclass{llncs}

\usepackage{balance} %
\usepackage{graphicx} %
\usepackage{times} %
\usepackage{url} %
\usepackage{verbatim} %
\usepackage{xcolor} 
\usepackage{color} 
\usepackage[small,labelfont=bf]{caption}

\usepackage{paralist}
\usepackage{indentfirst}
\makeatletter
\def\url@leostyle{%
\@ifundefined{selectfont}{\def\UrlFont{\sf}}{\def\UrlFont{\small}}}
\makeatother
\makeatletter
\let\@copyrightspace\relax
\makeatother

\newcommand{\yesfig}{true} %
\newcommand{\doFig}[1] {\ifthenelse{\equal{\yesfig}{true}}{{#1}}{}}
\definecolor{darkblue}{RGB}{0,0,105}
\definecolor{darkgreen}{RGB}{47,79,79}
\usepackage[colorlinks=true,linkcolor=darkblue,urlcolor=darkblue,citecolor=darkblue]{hyperref}

\definecolor{darkblue}{rgb}{0,0,0.45}
\usepackage{hyperref}

\begin{document}

\title{Bootstrapping Trust in Online Dating:\\ Social Verification of Online Dating Profiles}%

\author{Gregory Norcie$^{1,}$\thanks{Work done while the author was an intern at PARC.} \and Emiliano De Cristofaro$^2$ \and Victoria Bellotti$^2$}
\institute{$^1$ Indiana University $\;\;^2$ PARC (a Xerox Company)}

\maketitle

\begin{abstract}
Online dating is an increasingly thriving business which boasts billion-dollar revenues and attracts users in the tens of millions.
Notwithstanding its popularity, online dating is not impervious to worrisome trust and privacy concerns raised by the disclosure of potentially sensitive data as well as the exposure to self-reported (and thus potentially misrepresented) information. Nonetheless, little research has, thus far, focused on how to enhance privacy and trustworthiness. 
In this paper, we report on a series of semi-structured interviews involving 20 participants, and show that users are significantly concerned with the veracity of online dating profiles. To address some of these concerns, we present the user-centered design of an interface, called Certifeye, which aims to bootstrap trust in online dating profiles using existing social network data. Certifeye verifies that the information users report on their online dating profile (e.g., age, relationship status, and/or photos) matches that displayed on their own Facebook profile. 
Finally, we present the results of a 161-user Mechanical Turk study 
assessing whether our veracity-enhancing interface successfully reduced concerns in online dating users and find a statistically significant trust increase. 
\end{abstract}

\section{Introduction}
In the last few years, social networking has remarkably altered our social ecosystem. {\em Online Social Networks} (OSNs) offer highly efficient means for establishing or maintaining social connections. A 2011 survey of 6,000 people found that about 35\% of respondents reported spending more time socializing online than face-to-face and that 33\% of users were more likely to speak to someone new online than offline~\cite{badoo2012}.

Alas, the resulting ubiquitous gathering and dissemination of personal information also prompts some important privacy and trust concerns. The research community has begun to investigate how publicly sharing some kinds of personal information can help malicious entities launch various schemes such as creating personalized phishing attacks~\cite{jagatic2007social} or guess Social Security numbers~\cite{acquisti2009predicting}. 

Motivated by the significance and increasing recognition of trust and privacy issues~\cite{nielsen}, researchers have applied principles from  human-computer interaction to the design of security software, e.g., for social network chats~\cite{fahl2012helping}, anonymizing social graphs~\cite{felt2008privacy}, and file/email encryption~\cite{whitten1999johnny}. 
One of the most difficult challenges in enhancing trust, security, privacy in most networked systems has always been, and still remains, the human factor~\cite{huang2007survey,sasse2001transforming}. Awareness, perceptions, and reactions of non-tech-savvy users are subjective, and security experts and researchers have often struggled to gain understanding of the concerns of everyday users %
for whom security is not a primary task~\cite{dourish2004security}. Making this situation more difficult to contend with, study volunteers have a tendency towards social desirability distortion, wishing to present themselves in a favorable light to the experimenter by omitting information %
or even misrepresenting their behaviors~\cite{richman1999meta}. Thus, it is not surprising that participants report safer practices and higher privacy concerns than they actually demonstrate in practice~\cite{acquisti2006imagined}. 

As a result, we have been inspired to develop means for end users of social media to side-step awkward security and privacy mechanisms in achieving their social objectives. %
In particular, we focus on {\em Online Dating Services} (ODSs), attracted by their growing popularity and distinctive anthropological characteristics.
ODSs differ from typical OSNs, as they are almost exclusively used to connect with strangers. boyd~\cite{boyd2007social} defines OSNs as services allowing users to create and view web profiles, as well as to search other profiles to connect and communicate with. boyd also mentons that these connections 
almost exclusively involve existing or shared social contacts -- \textit{not} strangers. Conversely, the typical ODS \textit{does} aim to connect strangers.

Online dating has become remarkably popular: 20\% of heterosexual couples and 60\% of same-sex couples now report having met online~\cite{rosenfeld2012searching}. In 2012, more than 40 million users were estimated to be part of a \$1.9 billion revenue industry~\cite{stats}. Notwithstanding its popularity, online dating raises some worrisome privacy and trust issues. Because of the inherent need to engage with and reveal potentially sensitive information to unknown others, ODSs amplify many of traditional social-networking security and privacy issues. And yet, somewhat surprisingly, very little research has focused on the problem of how to make them more trustworthy and privacy-respecting. 

Motivated by the above concerns, this paper explores a number of potential ways to enhance trust and privacy in ODSs. 
We seek ODS users' input by conducting a series of semi-structured interviews (involving 20 participants) and concentrate on concerns that are particularly relevant to non-security-savvy users. We found that participants were particularly worried about the veracity of ODS profiles.
Following these semi-structured interviews, we present a user-centered design of a mock-up interface, called {\em Certifeye}, that allows users to certify some attributes, e.g., age, relationship status, photos, in their ODS profile. Certifeye does so by attesting that the information reported on the ODS profile matches that same information on the user's Facebook profile. Certifeye can be plugged on any existing ODS to add profile certification and bootstrap trust among users. In theory, profiles can be certified using a number of sources; however, driven by our user-centered design approach, and following a series of semi-structured interviews, we choose to use existing social network data for this purpose.
For instance, if a user is listed as ``in a relationship'' on Facebook, it would be cumbersome to change her status to ``single'', as friends and, most embarrassingly, the user's partner would likely notice this change.
Profile certification is performed seamlessly, by granting the ODS provider access to one's Facebook profile through Facebook API, which results into obtaining a ``certification badge''. Certifeye does not require users to mutually ``open'' their Facebook profiles and does not aim to replace the ODS credentials with Facebook credentials in the vein of OAuth~\cite{oauth} or OpenID~\cite{openid} technologies -- we only need a mechanism to verify the matching of information without accessing/storing any private information.

In order to assess whether or not a mock-up of our veracity-enhancing capability can successfully reduce trust concerns, we conducted a Mechanical Turk study involving 161 participants. The ODS users were asked to rank their concern on a scale from 1 to 7 (1 being not at all concerned, 7 extremely concerned), and we found a statistically significant increase in trust present when users were shown our proposed interface. %

\section{Related Work}
This section reviews prior work on trust and privacy in Online Dating Services (ODSs).
We also survey previous efforts toward the design of usable, private, and trustworthy Online Social Networks (OSNs) and analyze whether solutions applicable to OSNs can be adopted to ODSs.

\subsection{Trust \& Privacy Concerns in Online Dating}
Misrepresentation in OSNs has been recently investigated by Sirivianos et al.~\cite{sirivianos2012assessing}, who 
introduced ``FaceTrust'', a system using social tagging games to build assertion validity scores for profile information. 
Likewise, misrepresentation is an issue for ODSs as well and was first analyzed by Brym et al.~\cite{brym2001love}, who reported that 89\% of participants (ODS users) felt that ``people online might not tell you the truth about themselves'' and 85\% agreed that ``people you meet online might be hiding something.''
Ellison et al.~\cite{ellison2006managing} also pointed out that online daters must balance two conflicting goals -- they need to present oneself in the most positive light in order to attract a mate, while simultaneously knowing that one must be honest if one wants their relationship to progress past the first meeting. Thus, ODS users must balance positive self-presentation with transparency.
However, while ODS users might not lie about crucial traits, they do alter attributes that they consider minor, such as age or height. Toma et al.~\cite{toma2008separating} reported consistent, conscious misrepresentation. In their study, eighty participants were asked to recall the height, weight, and age listed on their ODS profiles. Participants were generally able to accurately recall the information on their ODS profile. The information on the participants profiles was found to be significantly different from what was reported on their driver's licenses. The deception was found to be \textit{not} subconscious, since participants were able to recall the false values. 

Recent research also shows that online daters daters take action when they suspect misrepresentation. Gibbs, Ellison, and Lai~\cite{gibbs2011first} found %
that many participants %
often engaged in information seeking activities, such as ``Googling'' a potential date.

In addition to trust issues, there seems to be some privacy concerns associated with using ODSs, including disclosing one's presence on a dating site. Couch et al.~\cite{couch2007online} described how the risk of ``exposure'' -- i.e., a coworker or acquaintance stumbling across one's profile. However, Couch's participants who reported exposure concerns were users of specialty fetish sites and/or sites geared specifically towards extremely short-term relationships. Conversely, our interviews focused on how users looked for medium to long term relationships, and we feel there is no longer a social bias against users seeking long term, monogamous relationships using ODSs.

Finally, Motahari et al.~\cite{motahari2009identity} discussed the concept of {\em social inference}, finding that that 11\% of study participants could correctly identify who they were communicating with, %
by utilizing out-of-band knowledge. For instance, someone may know there is only one female, hispanic local soccer team member, and use that outside knowledge to de-anonymize a seemingly anonymous profile. 

\subsection{Usable Trust \& Privacy in Online Social Networks}
\label{rwusable}
As we discussed earlier, although ODSs and OSNs share several similar traits (e.g., the ability of viewing profiles, listing interests, and exchanging messages), they actually provide different types of services and incur different challenges. Nonetheless, by examining prior work on OSNs, we aim to derive best practices for ODSs.

Privacy, trust, and security issues are often associated with the collection, retention, and sharing of personal information. One reason privacy concerns are pervasive in OSNs is because security is not a primary task. As Dourish et al.~\cite{dourish2004security} pointed out, users often view security as a barrier preventing them from accomplishing their goals. Furthermore, users may be unaware of the risks associated with sharing personal information. Data posted on social networks can be subject to subpoena or, even after years, can regrettably re-surface, e.g., during job hunting or an electoral campaign. Furthermore, social networking data can be used for social engineering scams. For instance, Jagatic et al.~\cite{jagatic2007social} showed that extremely effective phishing messages could be constructed by data mining social networking profiles to personalize phishing messages. %

Motivated by the significance of associated threats, a considerable amount of work has been dedicated to user-centered design of privacy and trust enhanced OSNs.
Privacy and trust are similar, but separate concepts. Nissenbaum~\cite{nissenbaum1998protecting} discussed the concept of ``contextual integrity'', pointing out that personal information is not simply private or public -- privacy depends on context.

As pointed out by Camp~\cite{camp2003design}, 
{\em trust} is separated from {\em privacy}, in that trust is the belief in the integrity or authority of the party being trusted. Thus, trust is extremely closely connected to \textit{veracity} and \textit{reputation}.
Nonetheless, trust is similar to privacy in that it must be incorporated into software's design. Naturally, the task of effectively incorporating privacy and trust into a design may be challenging. Cavoukian~\cite{cavoukian2010privacy} described 7 principles of ``privacy by design.'' These principles aim to embed privacy and data protection a throughout the entire life cycle of technologies, from the early design stage to their deployment, use and ultimate disposal. 

Similarly, Murayama et al.~\cite{murayama2006anshin} discussed how the Japanese concept of ``\textit{anshin}'' -- the emotional component of trust -- can be taken into account when building systems. While the concept of \textit{anshin} may not be known by name to westerners, the concept itself is not new. For example,  Bruce Schneier used the term  ``security theater''~\cite{schneier2003beyond} to describe security measures taken by the TSA to increase the public's feeling of trust in flying post 9/11 which serve no useful purpose.

It could be argued that acts of security theater are attempts to create \textit{anshin}. By operationalizing \textit{anshin} in this manner, we can see that it is important to increase trust in a system, as well as see that failed attempts to increase trust can lead to user frustration.
 
Another issue typical of OSN is {\em over-sharing}. When social networks do not embed privacy into their designs, users tend to over-share and make dangerous errors. For instance, Wang et al.~\cite{wang2011regretted}
surveyed 569 Facebook users and found that 21\% of users had regretted posting information on Facebook, with regrets usually centering around sensitive content, strong sentiments, or because the post exposed a lie or secret. Moreover, even content that users do not regret posting can have privacy implications.
Gross and Acquisti~\cite{gross2005information} crawled the profiles of Carnegie Mellon University's Facebook population in 2005, and found that 90.8\% of profiles publicly displayed images, 39.9\% publicly displayed phone numbers, and 50.8\% publicly displayed their current residence. 
They also found that most users had not changed their privacy settings from Facebook's defaults. 
Sharing this kind of information can be harmful, aiding an attacker in various re-identification attacks, such as guessing a user's Social Security numbers based on publicly available information~\cite{acquisti2009predicting}.%

In summary, while prior work has focused on privacy and trust in OSNs, or analyzed misrepresentation in ODSs, our work is the first, to the best of our knowledge, to present a user-driven and user-centered design of an ODS interface that enhances trust by leveraging information that is already available in OSNs.

\section{Study Part 1: Ideation and Interviews}\label{sec:part1}
As mentioned in Section I, a remarkable amount of people who classify themselves as ``single and looking'' use Online Dating Services (ODSs) today. Naturally, online dating presents numerous privacy and trust issues, yet, little work has focused on them.
In order to avoid unnecessary effort on unfocused interviews (e.g., covering {\em all} possible concerns about online dating), our first step was to brainstorm on a few concepts for privacy and trust enhanced online dating, informed by what prior research suggests as promising problem areas. Two security and privacy researchers held a series of informal brainstorming sessions and then reviewed and evolved the ideas with a third researcher experienced in user-centered design. 

\subsection{Initial Concepts}

Based on our expert brainstorming session, we came up with four possible privacy and/or trust enhancements to existing ODSs.
\begin{enumerate}
\item \textbf{\em Identifying potential romantic partners in social circle without trusted third parties:}
Consider the following scenario: Alice is attracted to Bob, but she does not want to reveal her sentiment, unless Bob also likes Alice. 
Alice belongs to a social networking service (e.g., Facebook), and installs an application that lets her list the people that she would like to date. By utilizing appropriate privacy-enhancing technologies (e.g.,~\cite{de2010practical}), this information could be exchanged and stored in such a way that: (i) users only learn whether there is a match, and (ii) the provider does not obtain any information about users' interests and/or matches.
\item \textbf{\em Identifying potential  partners based on matching interests, without trusted third parties:} Many ODSs ask extremely personal questions, the answers to which users would like to reveal only to prospective romantic partners. Similar to the above idea, users' answers could be exchanged and stored in a privacy-preserving manner, so
that only users with a minimum number of matching interests would disclose personal information.
\item \textbf{\em Automatic Exclusion of Coworkers/Friends:} It may often be embarrassing to admit to friends that one is using ODSs to find someone for a certain kind of relationship. Therefore, it might be useful to grant the dating service access to their list of Facebook friends, and exclude friends, coworkers, and/or family members from seeing their profile.
\item \textbf{\em Certification of Profile Data:} A natural concern in ODSs is related to veracity of user profiles. It seems possible that some service could pull data from another social source (e.g., from Facebook) and certify on the dating site that the information is accurate. 
\end{enumerate}
Having developed these four ideas, our next step was to determine which of them (if any) matched the users' needs. Therefore, we conducted a series of semi-structured interviews where we asked users to describe their Facebook habits and their ODS habits. We also posed several questions specifically designed to explore our initial ideas. (For example, would users be willing to link their social networking profile to their ODS profile?) We discuss these interviews in details below.

\subsection{Interview Methodology}
We recruited 20 users from a local classified advertisements website, as well as from mailing lists at a local university.\footnote{Our studies obtained the Exempt Registration status from PARC's Institutional Review Board (FWA Number: FWA00018829, Expiration 5/2/2017).} Interviewees were required to be past or present users of ODSs. The male to female ratio was roughly equal (55\% female). Participants ranged in ages from 24 to 70 and were mostly educated, with 75\% of users possessing at least a bachelor's degree. Age and education breakdowns are reported in Table~\ref{age-int}
and \ref{edu-int}.

\begin{table}[tbp]
\centering
\parbox[t][][t]{.45\linewidth}{\centering
\centering
{\renewcommand{\arraystretch}{1.05}
\vspace{-0.93cm}
\begin{tabular}{| l | c | r |}
\hline
{\bf Age} & {\bf N} & {\bf \%} \\
\hline
18--25 & 6 & 30\%\\
26--35 & 4 & 20\%\\
36--55 & 8 & 40\%\\
55--70 & 2 & 10\%\\
\hline
\end{tabular} 
\vspace{0.15cm}
\caption{Age breakdown for interviewees}
\label{age-int}}
}
\hfill
\centering
\parbox[t][][t]{.5\linewidth}{\centering
{\renewcommand{\arraystretch}{1.05}
\begin{tabular}{| l | c | r |}
\hline
{\bf Degree} & {\bf N} & {\bf \%}\\ \hline
Some college, no degree & 4 & 20\%\\
Associate's & 1 & 5\%\\
Bachelor's & 6 & 30\%\\
Master's & 8 & 40\%\\
PhD & 1 & 5\%\\
\hline
\end{tabular}
\vspace{0.15cm}
\caption{Education breakdown for interviewees}
\label{edu-int}
}
}
\vspace{-0.4cm}
\end{table}

Prior to the interview, participants were asked to fill out a short online survey. Besides demographic information, they were presented with some multiple-choice question covering their ODS habits, the services they use, the types of information they post on Facebook, and whether or not they refrain from posting certain types of information on Facebook. Participants were also asked about where they meet partners offline. 
After the computer survey was filled out, a semi-structured face-to-face interview session was arranged where users were asked to log into their Facebook account. We then asked the interviewee to scroll down their ``feed'' until he or she came across an embarrassing, controversial, or ``edgy'' post, either by themselves or others. Such ``provocative'' items were used to drive the discussion, aiming to tap into our interviewees real responses to social actions (e.g., disclosures within the social networking environment), rather than have them imagine how they might feel about abstract privacy and trust issues. While examining the mini-feed, users were asked questions such as:
\begin{compactitem}
\item Why did you choose to post this?
\item Who can see this post? Did you consider that when posting this?
\item Your friend made this update: would you post something like this? Why/why not?
\end{compactitem}
We then asked the interviewees to discuss whether or not they would be willing to use a theoretical online dating application which existed within Facebook, and whether they would be willing to share their Facebook information with an ODS. The total time spent on the computer survey and interview was, on average, approximately 45 minutes.

\subsection{Interview Results}
We anticipated that a relevant concern for our interviewees would be related to disclosing, e.g., to their social circle, the fact that they dated online. Previous work~\cite{couch2007online} had found that some ODS users feared exposure of their ODS habits, however, this concern was limited to users of ``an online dating website focused on sexual interests'', whereas, our interviews were focused on how interviewees used more typical romance-oriented ODSs -- with fewer potentially embarrassing connotations. 
We found that a majority of interviewees (15/20) did not see online dating as very embarrassing.
While many participants (14/20) did not want their entire social graph (including loose ties such as coworkers and/or acquaintances) knowing they used online dating, these interviewees were fine with close friends and family knowing that they used an ODS.

However, the participants still had some concerns. 75\% of interviewees (15/20) did not want their mini-feeds to reflect their use of dating apps. Further, 25\% of users expressed a belief that using an online dating app on Facebook would mix social circles in an undesirable fashion.
While some participants did not wish to mix social circles and date within their social network (4/20), other users simultaneously complained that online dating ``didn't work' (3/20) or that ``chemistry is more than a profile'' (7/20). Nonetheless, the users did not mind if their close friends or acquaintances saw them on an ODS. One interviewee compared being seen on an ODS to being seen at a gay bar: {\em ``They can't really judge me, cause, hey, they're here too!''} Thus, we conclude that users were not particularly concerned with excluding coworkers and/or friends.

One problem that users did express was related to the veracity of ODS profiles. Users overwhelmingly felt that they could not trust that the data they found in ODSs was accurate. Older users were concerned that that profiles misrepresented age and/or relationship status, while younger participants were more concerned that the photos posted were either altered, out of date, or taken from an especially flattering angle (often referred to as the ``MySpace Angle''.)

Since we had initially considered creating a privacy-preserving dating application for Facebook, our pre-interview survey asked users if there was any information the interviewee refrained from posting to Facebook, then followed up with a question asking ``What security measures would Facebook have to take for you to be willing to share this information?'' All participants indicated that they had such information and about half (11/20) responded that they ``simply refuse to post this information'' and that no technical measure could change their minds. In the interviews, participants also revealed that certain information, such as address and/or phone number, was simply too private to be entrusted to Facebook.
Overall, our analysis clearly showed that concerns about information disclosure were actually less relevant to users than
the issue of veracity of information contained in profiles. 
Interestingly, all participants agreed that a Facebook profile contained enough information to make a decision about whether to date someone. This suggested that Facebook would be an ideal source of information to bootstrap trust in ODS.

{\em Note: Our analysis of the interviews focused on discovering common themes and complaints to drive the design of our application (which was tested in a rigorous manner, as noted later). Thus, any observations above do not necessarily extend to the general population.}

\begin{figure}[t]
\vspace{0.1cm}
\centering
\includegraphics[width=0.45\linewidth]{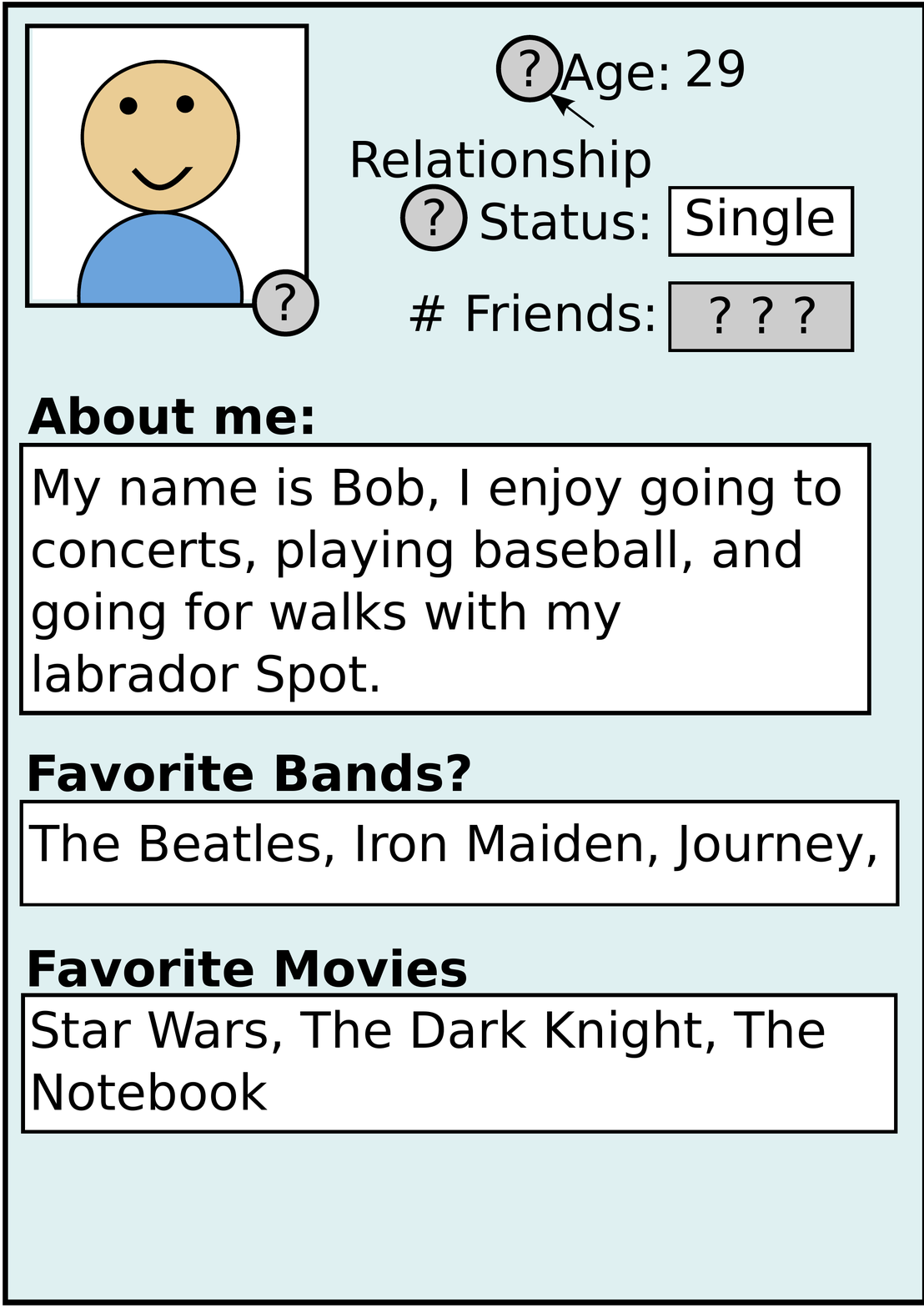} 
\vspace{-0.2cm}
\caption{An example profile shown to users in our study.}
\label{fig:profile}
\vspace{-0.3cm}
\end{figure}

\section{Study Part 2: The Certifeye Interface}
Based on our interviews, we found that the direction of certifying profile data direction would be the most valuable to ODS users. We called our system concept ``Certifeye'' and designed it as a Facebook and ODS application allowing users to certify that their ODS profile information was accurate.
Aiming to address the problem of doubtful profile veracity, Certifeye users could pull their relationship status, age, and photos from their Facebook account, and receive green badges to show that this information had been certified. 

\begin{figure}[t]
\centering
\includegraphics[width=0.52\linewidth]{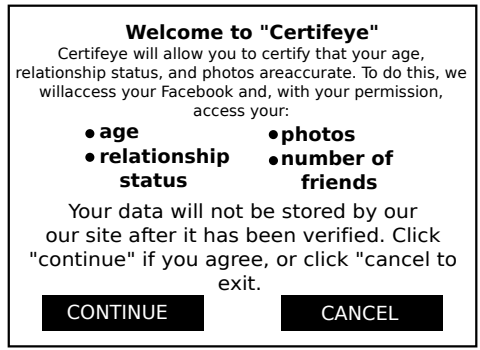} %
\vspace{-0.2cm}
\caption{The Certifeye consent screen.}
\label{fig:consent}
\vspace{-0.25cm}
\end{figure}
\begin{figure}[b]
\vspace{-0.15cm}
\centering
\includegraphics[width=0.52\linewidth]{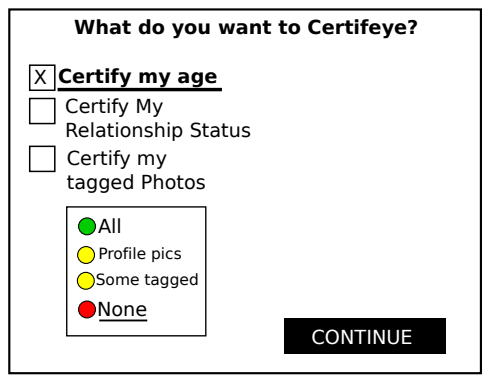} %
\vspace{-0.25cm}
\caption{The main Certifeye interface}
\vspace{-0.5cm}
\label{fig:ui}
\end{figure}

We developed a visual mock-up of Certifeye for use in end-user evaluations. The interface consists of three main screens. First, users are greeted with a generic ODS profile (Figure~\ref{fig:profile}) containing various personal information items such as profile photo, age and relationship status, each with a certification badge displayed next to it. At the outset, all certification badges are greyed out and display question marks.%
Upon clicking on a certification badge, Certifeye displays a consent screen (Figure \ref{fig:consent}). It should be noted that the Certifeye UI also displays the numerical total of a user's Facebook friends. As discussed in Section~\ref{sec:discussion}, this is an important feature, since fake profiles will be harder to create if they must have large numbers of friends. Creating fake profiles requires technical expertise, time, and cost -- barriers that may push dishonest users to other, less trustworthy ODS services.

Finally, after consenting to the syncing of her Facebook account with her ODS profile, the user is presented with the main Certifeye interface, where (s)he can choose to certify relationship status, age, and/or photos (Figure~\ref{fig:ui}). 
Observe that we embed trust in our design by envisioning periodic user interaction to ``renew'' the certification. After a user has gone through the Certifeye interface once, badges are not updated when her Facebook information changes. Therefore, after a set period of time, the badges revert to grey question marks. The Facebook API would actually grants ongoing access to a user's profile information once a user initially consents, however, we prefer to respect users' privacy and let them explicitly consent to a new information access.

\subsection{The Turk Experiment}
After creating our prototype interface, we then showed this interface to 161 Mechanical Turk users. Users were asked if they were current ODS users, and users who were not ODS users were not allowed to proceed with the survey.
Users who were not screened out were asked to rank their concern with the veracity of several types of ODS information with and without our trust enhancements.
Participants were even more representative of the general population than those in our initial interviews.
(Age and education breakdowns are reported in Table~\ref{tab:ageTurk}
and \ref{tab:eduTurk}.)
Also, remind that Mechanical Turk workers
have been shown to be as reliable as traditional participant pools for human subject research~\cite{paolacci2010running}. 
\begin{table}[t]
\centering
\parbox[t][][t]{.4\linewidth}{\centering
\centering
{\renewcommand{\arraystretch}{1.05}
\vspace{-0.53cm}
{\renewcommand{\arraystretch}{1.05}
\begin{tabular}{| l | c | r |}
\hline
{\bf Age} & {\bf N} & {\bf \%} \\ \hline
18--22 &19 & 11.8 \%\\
23--30 & 65 & 40.4\%\\
31--50 & 67 & 41.6\%\\
51--70 & 20 & 6.2\%\\
\hline
\end{tabular}
}
\vspace{0.15cm}
\caption{Age breakdown for Turkers}
\label{tab:ageTurk}}
}
\hfill
\centering
\parbox[t][][t]{.52\linewidth}{\centering
{\renewcommand{\arraystretch}{1.05}
\begin{tabular}{| l | c | r |}
\hline
{\bf Education Level} & {\bf N} & {\bf \%} \\ \hline
9th -- 12th, no diploma &2 &1.2\%\\
HS (includes GED) &27 &16.8\%\\
Some college (no degree) &43 &26.7\%\\
Associates &25 &15.5\%\\
Bachelor's Degree &40 &24.8\%\\
Master's Degree &20 &12.4\%\\
PhD &4 &2.5\%\\
\hline
\end{tabular}
\vspace{0.15cm}
\caption{Education breakdown for Turkers}
\label{tab:eduTurk}
}
}
\vspace{-0.5cm}
\end{table}

Participants were asked for basic demographic information, then asked to rate their comfort with three scenarios on a scale from 1 to 7 (1 = Not at all concerned, 7 = Extremely concerned), in response to the following questions:

\begin{compactenum}
\item How concerned are you with people misrepresenting their relationship status on online dating sites?
\item How concerned are you about people misrepresenting their age on online dating sites?
\item How concerned are you about users on online dating sites misrepresenting themselves using old, altered, or engineered photos? (Examples: Photoshopped pictures, the ``MySpace Angle'')
\end{compactenum}

Participants were also asked other questions (e.g., ``How concerned are you about computer viruses?'') to avoid biasing the participants towards trying to give ``correct'' answers. The questions were also presented in random order to eliminate any positioning effects.
After participants had answered the questions, they were displayed a series of mocked-up screenshots from the Certifeye interface. Users were taken through as ``Bob'', a hypothetical user of Certifeye, exploring the functionality of the software via a series of annotated screenshots. Along the way, the functionality of Certifeye was revealed to Bob (and thus, to the participant).
After being shown the interface, Turkers were asked to rank how concerned they were with misrepresentation \textit{assuming an ODS used the interface they had just seen}.
As Kittur et al.~\cite{kittur2008crowdsourcing} pointed out, Mechanical Turkers often try to cheat at tasks. To guard against this, a series of short ``sanity-check'' questions ensured that users had paid attention to the interface. For example, a screen might state that Bob clicked on ``None'', and all the faces in the interface turned red. We would then ask the user ``What color did the faces in the interface turn?'' Users who did not answer all sanity checks correctly were not included in the analysis. Out of an initial pool of 200 users, 39 users failed their sanity check, leaving 161 valid responses for analysis.

\subsection{Turk Study Results}\label{sec:turkresults}
Since levels of comfort are not assumed to be normally distributed, we used a Mann-Whitney U test to check that the changes in average Likert scores presented in Table~\ref{tab:turkLikert} were statistically significant. We asked participants to rank on a scale from 1 to 7 how concerned they were that the ages, relationship statuses, and photos on an ODS which used Certifeye were accurate (1 = Not at all concerned, 7 = Extremely concerned). We found that users felt more comfortable with the Certifeye interface, 
and that the difference was statistically significant. Specifically, for, age, relationship status, and photos, respectively, Mann-Whitney U values are equal to 7978, 8326, and 7693.5. n1 = n2 = 161 and P $<$ 0.01 (2-tailed).
As 1 represented not concerned at all, and 7 extremely concerned, this means that new level of concern is positively below neutral (4/7), while the previous was not, thus, we conclude that the features provided by Certifeye---verification of age, photos, and relationship status---reduce users' concerns with respect to information's veracity. 

\begin{table}[h]
\vspace{-0.65cm}
\centering
\bigskip
{\renewcommand{\arraystretch}{1.05}
\begin{tabular}{| l | c | c | r |}
\hline
{\bf Type of Info} & {\bf Concern Pre} & {\bf Concern Post} & {\bf P-Value} \\ \hline
Photos & 4.7 & 3.5 & $<.001$\\
Rel. Status & 4.6 & 3.4 & $<.001$\\
Age & 4.1 & 3.1 & $<.001$\\
\hline
\end{tabular}
}
\vspace{0.1cm}
\caption{Results of Mechanical Turk Likert questions.}
\label{tab:turkLikert}
\vspace{-0.65cm}
\end{table}

In our Mechanical Turk study, before being shown our interface, users were asked to rank some criteria as for how important they are when deciding if a stranger's Facebook profile is genuine. (1 = Most important, 8 = Least important.) The order of options was randomized for each participant to avoid biasing respondents to any particular item. Below, we report suggested criteria, ordered from most important to least important (according to participants in our study):

\begin{compactenum}
\item Number of friends
\item Location (city)
\item Workplace
\item College / Grad School 
\item Mutual interests
\item Attractiveness / appearance
\item Other
\end{compactenum}

As per ``other'', common responses included having mutual friends, and whether the profile appeared ``spammy''. 
As a result, we conclude that ``Number of friends'' was the most preferred criteria to assess whether or not a social network profile was genuine.

\section{Discussion}\label{sec:discussion}
As mentioned earlier, our exploration of privacy and trust in ODSs has been user-driven. We interviewed users based on a few preliminary concepts, such as, (1) Identifying potential romantic partners within one's social circle without trusted third parties, 
(2) Listing sensitive information without a trusted third party, (3) Automatic exclusion of coworkers/friends, and (4) Certification of profile data.

While we only pursed the certification of profile data (motivated by a stronger interest of users), we now report on some lessons learned during our interviews.
As mentioned in Section~\ref{sec:part1}, we asked users whether they would like an application identifying potential romantic partners, both (within and outside one's social circle), without the need to disclose ``private'', possibly sensitive, information to any third party OSN or ODS sites. While such application would naturally enhance users' privacy, we found that users did not see the point of using a computer system to meet people they knew in real life. Thus, while interviewees were comfortable with linking Facebook profiles to their ODS accounts, they did not see the point in doing so. For instance, one user reported {\em ``If I want to ask someone out, I'll ask them out -- I don't need a computer to do that''}. Even when prompted as to whether being connected with ``friends of friends'' would be a useful function, interviewees were dubious as their friends could just introduce them to potential matches.
In general, interviewees seemed reasonably comfortable with online dating but agreed that they did not want their online dating information sent to the general public. Some participants specifically mentioned that they would not use a dating application that posted information to their Facebook mini-feed.

As mentioned in the interview results, users did not mind if their friends or acquaintances knew they used an ODS. Also, even if users did find the idea of automatic exclusion of coworkers/friends useful, terms of service of most social network sites, e.g., Facebook, do not actually allow APIs to access social graph data of users who have not opted in (therefore, this approach would be infeasible).
We also found that, while misrepresentation is a widespread concern, the specific kind of information actually concerning users varied with age. Older users were concerned about misrepresentation of age and/or relationship status. Relatively older women were especially wary that older males may misrepresent their marital status, with many citing bad experiences which led to such distrust. However, regardless of gender, older users wanted to verify that age and relationship status were accurate, whereas, younger users were more concerned about accurate, recent photos being present on the ODS. 

As a result, trust being a major concern, we decided to explore the user-centered and usable design of mechanisms to bootstrap trust in ODS, by leveraging social-network based reputation. This approach presented several interesting challenges. 
First, a user could set up a fake social network profile, link it to her ODS profile, and ``Certifeye'' it despite the fact that her information is actually false. As mentioned in Section~\ref{sec:turkresults}, study participants listed ``number of friends'' as the most preferred criteria to assess whether a stranger's Facebook profile was genuine or fake. So, we designed our software to always include the number of friends along with the certification data. Specifically, whenever a user certifies her profile, in addition to green or red badges, the profile is also populated with a small box listing how many friends she has.
Also, note that Facebook actively takes steps to detect and remove accounts that it deems to be secondary or fake. This makes it hard for ordinary users to maintain fake profiles long enough to gain a substantial number of friends (Facebook looks for suspicious patterns and blocks suspect accounts but, unsurprisingly, will not reveal how it does this~\cite{facebook2010}.)

Although users may friend unfamiliar people~\cite{ryan2010getting}, Pew~\cite{pew2012} has shown that most Facebook users have an average of 229 friends. Spammers (i.e., malicious entities with an economic incentive to abuse the system) tend to have more friends than average~\cite{zdnet2012}, however, this requires a non-trivial time and economic commitments that are unrealistic for ordinary people only trying to make their dating profile more appealing. Also, creating fake profiles also increases the risk of the profile being blocked by Facebook. Thus, these barriers will arguably push dishonest users to other less trustworthy ODSs. Previous work -- for instance, in the context of spam~\cite{liu2006proof} -- has showed that raising costs for malicious users causes them to move to easier-to-abuse services.

Nonetheless, if we were to develop an actual service based on the Certifeye design, we would further enhance trust by also taking into account additional information, such as date of most recent status update or other automated fake account detection mechanisms, such as, the one proposed by Sirivianos et al.~\cite{nsdi12}. Also, we would need provide the users with information about how to interpret what data they see in another profile. 

Certifeye also allowed the certification of ODS pictures. Users could earn a yellow badge by sharing all of their profile pictures, and a green badge by sharing all of their tagged photos. While, theoretically, users could untag themselves from unflattering photos before syncing their Facebook profile with our certification application, we feel this is not a limitation, as already pointed out in previous work. Facebook profiles serve as a form of self-presentation~\cite{boyd2007social}, thus, if Facebook users wish to share one facet of themselves with an ODS, and another facet of themselves with their Facebook friends, they would be forced to untag their unflattering photos, certify their profile, then retag themselves. Once again, raising the effort required to create fake profile data will likely prompt ``malicious'' users to move on to other, less well-protected ODSs.

Finally, while we describe our social verification based on Facebook profiles, note that our techniques are not limited to one specific OSN. We use Facebook as an example, motivated by its widespread penetration (900 million users in 2012~\cite{cnn2012}.) Nonetheless, our techniques could work with any ODS and with any OSN providing APIs to share age, relationship status, and pictures. We could easily modify our workflow so that Certifeye could interface with other services, such as Google Plus, Orkut, Diaspora, etc.

\section{Conclusion} %
This paper analyzed %
concerns of Online Dating Services (ODSs) users 
about the veracity of information presented in the profiles of potential dates. 
Motivated by the results of semi-structured interviews (involving 20 users),
we designed an interface, called Certifeye, that lets users certify some attributes, such as age, relationship status, and photos, in their ODS profile. Our prototype does so by attesting that the information reported on the ODS profile corresponds to that on the user's own Facebook profile. We ran a Mechanical Turk study with 161 users to assess whether or not a mock-up of this veracity-enhancing capability successfully reduced trust concerns and, indeed, we found a statistically significant reduction when users were presented with Certifeye.

Naturally, our work does not end here. 
We plan to develop our Certifeye interface and integrate it in an actual ODS.
Also, we intend to further explore privacy and trust concerns in the context of both ODSs and OSNs, and deploy usable privacy-enhancing technologies, following similar user-driven approaches.

\end{document}